\documentclass{ws-procs9x6}

\begin{document}

\title{NNLO analysis of unpolarized DIS Structure Functions}

\author{J.~BL\"UMLEIN, H.~B\"OTTCHER}

\address{DESY 
Platanenallee 6, 
15738 Zeuthen, Germany}

\author{A.~GUFFANTI}

\address{School of Physics,
University of Edinburgh  
King's Buildings, 
Mayfield Road, 
Edinburgh EH9 3JZ, United Kingdom}  

\maketitle

\abstracts{
We present the results of a NNLO QCD analysis of the World data
on unpolarized DIS Non-Singlet Structure functions.}

\vspace{-9cm}
\begin{flushright}
DESY 06-053\\
Edinburgh 2006/12
\end{flushright}
\vspace{7cm}
\section{Introduction}


The increasing accuracy of DIS experiments will further reduce the 
experimental errors on the determination of the strong coupling constant 
calling for an improvement on the theoretical errors, which are by now the 
dominant ones.  
One way to achieve it is to include NNLO QCD effects in the analysis.

Our goal is to perform a NNLO QCD analysis of World data on unpolarized DIS 
structure functions to determine $\alpha_s$ with an accuracy of ${\mathcal O}(2\%
)$ along with a parametrization of the parton distribution functions 
with fully correlated errors. As a first step in this direction we 
concentrate on the non-singlet (NS) sector.

We presented the first results of our analysis in\cite{Blumlein:2004ip}. 
Here we give an update of the main results and refer the interested 
reader to\cite{BBG} for all the details.
\vspace{-.1cm}
\section{Theoretical Framework}

We carried out our analysis in Mellin$-N$ space\cite{BV}, where the non-singlet
part of the electromagnetic DIS structure function $F_2(N,Q^2)$ is 
written in terms of the non-singlet quark combinations $q^{\pm,v}(N,Q^2)$ 
and the corresponding Wilson coefficients $C_2^{(k)}(N)$ as
\begin{equation}
F_2^{\pm,v}(N,Q^2)=\{1+a_s(Q^2) C_2^{(1)}(N) + a_s^2(Q^2)C_2^{(2)}(N)\}
                                                        q^{\pm,v}(N,Q^2)\,,
\end{equation}
with $a_s(Q^2)\equiv\alpha_s(Q^2)/4\pi$, the normalised coupling constant.

In the region $x>0.3$ we adopt the quark valence dominance hypothesis under 
which the proton and deuteron structure functions are given by the following
quark distribution combinations
\begin{equation}
  F_2^{p}=\frac{4}{9}xu_v+\frac{1}{9}xd_v\,,\qquad
  F_2^{d}=\frac{5}{18}x(u_v+d_v)\,.
\end{equation}
For $x<0.3$ we analyse the NS combination
\begin{equation}
  F_2^{NS}\equiv2(F_2^p-F_2^d)=\frac{1}{3}x(u_v-d_v)
                             -\frac{2}{3}x(\overline{d} -\overline{u})\,.
\end{equation}

The valence parton distribution functions are parametrized at the
reference scale $Q_0^2=4$ GeV$^2$ with the functional form
\begin{equation}
xu_v(Q_0^2,x)=A_u x^{a_u}(1-x)^{b_u}(1+\rho_u\sqrt x + \gamma_u x)
\end{equation}
and
\begin{equation}
xd_v(Q_0^2,x)=A_d x^{a_d}(1-x)^{b_d}(1+\rho_d\sqrt x + \gamma_d x)\,,
\end{equation}
where the normalization constants $A_u$ and $A_d$ are not free parameters
of the fit but are determined to satisfy the valence quark counting:
$\int_0^1u_v(x)dx=2$ and $\int_0^1d_v(x)dx=1$.

The remaining non-singlet parton density, $(\overline{d} -\overline{u})$,
is not constrained by the electomagnetic structure function data and we
adopted the form given in\cite{Martin:2001es},
which provides a good description of the Drell-Yan dimuon production data
from the E866 experiment. The heavy flavor corrections were accounted for
as described in\cite{AB}.
\vspace{-.25cm}
\section{Data}

The results we present are based on 551 data points for the structure 
function $F_2(x,Q^2)$ measured on proton and deuteron targets.
The experiments contributing to the statistics are: BCDMS\cite{Benvenuti:1989rh},
SLAC\cite{Whitlow:1991uw}, 
NMC\cite{Arneodo:1996qe}, H1\cite{Adloff:2000qk} and ZEUS\cite{Breitweg:1998dz}.

The BCDMS data were recalculated replacing $R_{QCD}$ with 
$R_{1998}$\cite{Abe:1998ym}. All deuteron data were corrected for Fermi motion 
and off-shell effects\cite{Melnitchouk:1995fc}.

We used the measured structure functions $F_2^p$ and $F_2^d$ in the 
region $x>0.3$ which is expected to valence dominated, while in the region 
$x<0.3$ we construct the non-singlet structure function 
$F_2^{NS}\equiv 2(F_2^p-F_2^d)$ from proton and deuteron data measured at 
the same $x$ and $Q^2$.

We imposed different cuts on the data. Only data points with
$Q^2>4$ GeV$^2$ were included in the analysis and a cut on the hadronic mass
of $W^2>12.5$ GeV$^2$ was imposed in order to reduce higher twist effects on
the determination of $\Lambda_{QCD}$ and the PDF parameters. The latter cut
was then relaxed in the extraction of higher twist effects.
Moreover we imposed additional cuts on BCDMS ($y_\mu>0.3$) and NMC 
($Q^2>8$ GeV$^2$) data in order to exclude regions with potentially 
significant correlated systematic errors.

In the fitting procedure we allowed for a relative normalization shift
between the different data sets within the systematic uncertainties
quoted by the single experiments. These normalization shifts were fitted
once and then kept fixed.
\vspace{-.2cm}
\section{Results}

The results we obtain for the fit parameters are collected in Table~\ref{tab1}.
We note that the fit doesn't constrain the $\rho_i$ and $\gamma_i$
parameters, which have therefore been kept fixed after the first minimization 
and their value is quoted without errors.

The remaining parameters to be determined in the fit are the low- and 
high-$x$ ones ($a_i$ and $b_i$) alongside with $\Lambda_{QCD}$.
\vspace{-.2cm}
\normalsize
\begin{table}[th]
\tbl{Parameters values determined in the NNLO QCD fit.}
{
\begin{tabular}{|c|c|c|}
\hline
$u_v$      & $a$         &  0.291 $\pm$ 0.008 \\
           & $b$         &  4.013 $\pm$ 0.037 \\
           & $\rho  $    &  6.227             \\
           & $\gamma$    & 35.629             \\
\hline
$d_v$      & $a$         &  0.488 $\pm$ 0.033 \\
           & $b$         &  5.878 $\pm$ 0.239 \\
           & $\rho  $    & -3.639             \\
           & $\gamma$    & 16.445             \\
\hline
\multicolumn{2}{|c|}{$\Lambda_{QCD}^{(4)}$, MeV} & 226 $\pm$ 25 \\
\hline \hline
\multicolumn{2}{|c|}{$\chi^2 / ndf$} & 472/546 = 0.86 \\
\hline
\end{tabular}\label{tab1}}
\end{table}
\vspace{-.2cm}
From the value for $\Lambda_{QCD}^{(4)}$ obtained in the fit we 
extract the following value for the strong coupling constant
\begin{equation}
\alpha_s(M_Z^2) = 0.1134\;
                \begin{array}{r}+0.0019\\-0.0021\end{array}\;\;\;
		\mathrm{(expt)}\,. 
\end{equation}
We note that this value is in agreement within the errors with results
obtained from other NNLO QCD analyses\cite{Martin:2004ir,Alekhin:2005gq} 
and with the the world average $0.1182\pm0.0027$\cite{Bethke:2004uy} within $2 \sigma$.

\begin{figure}
\begin{center}
\includegraphics[width=5cm]{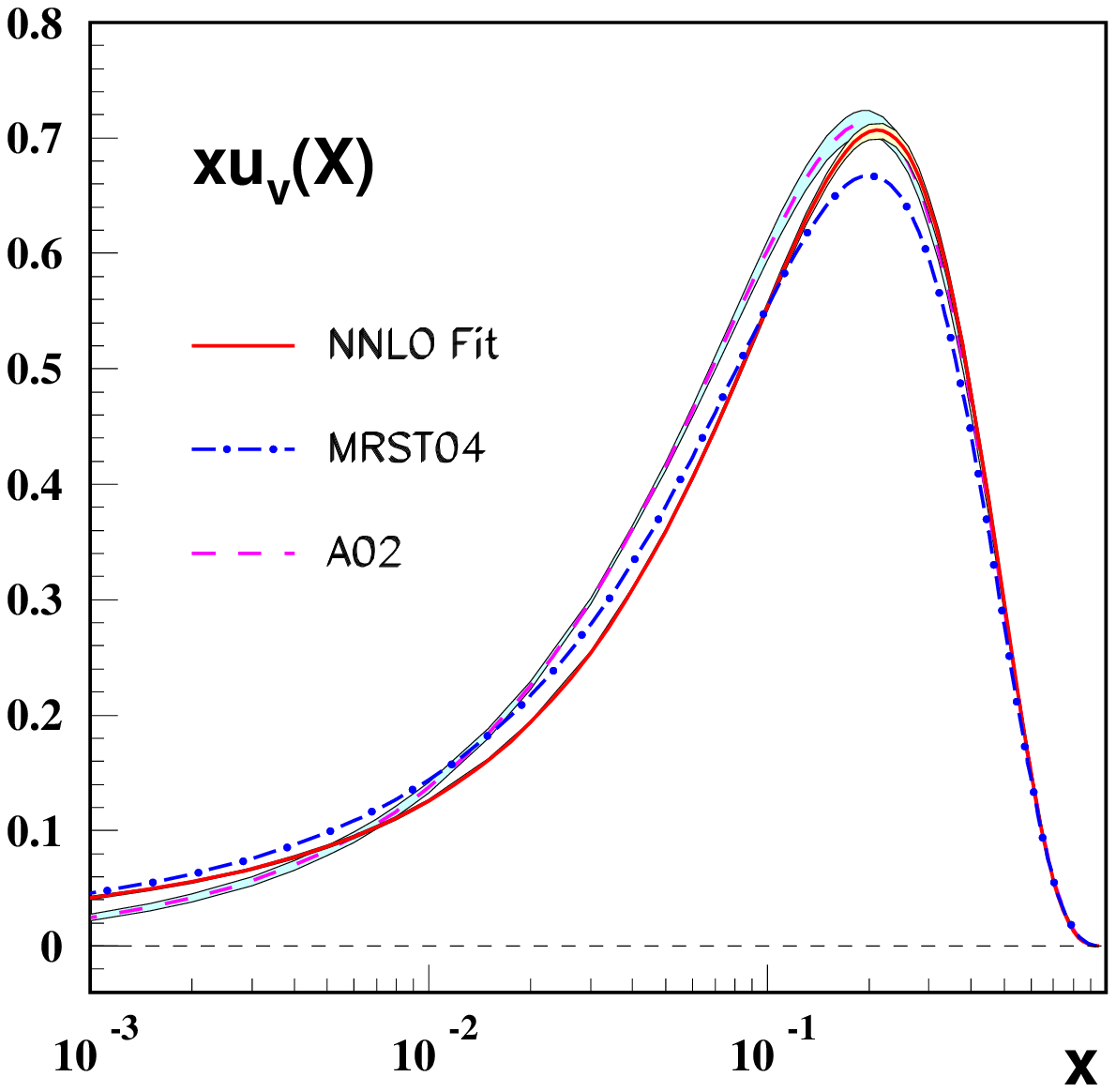}\qquad
\includegraphics[width=5cm]{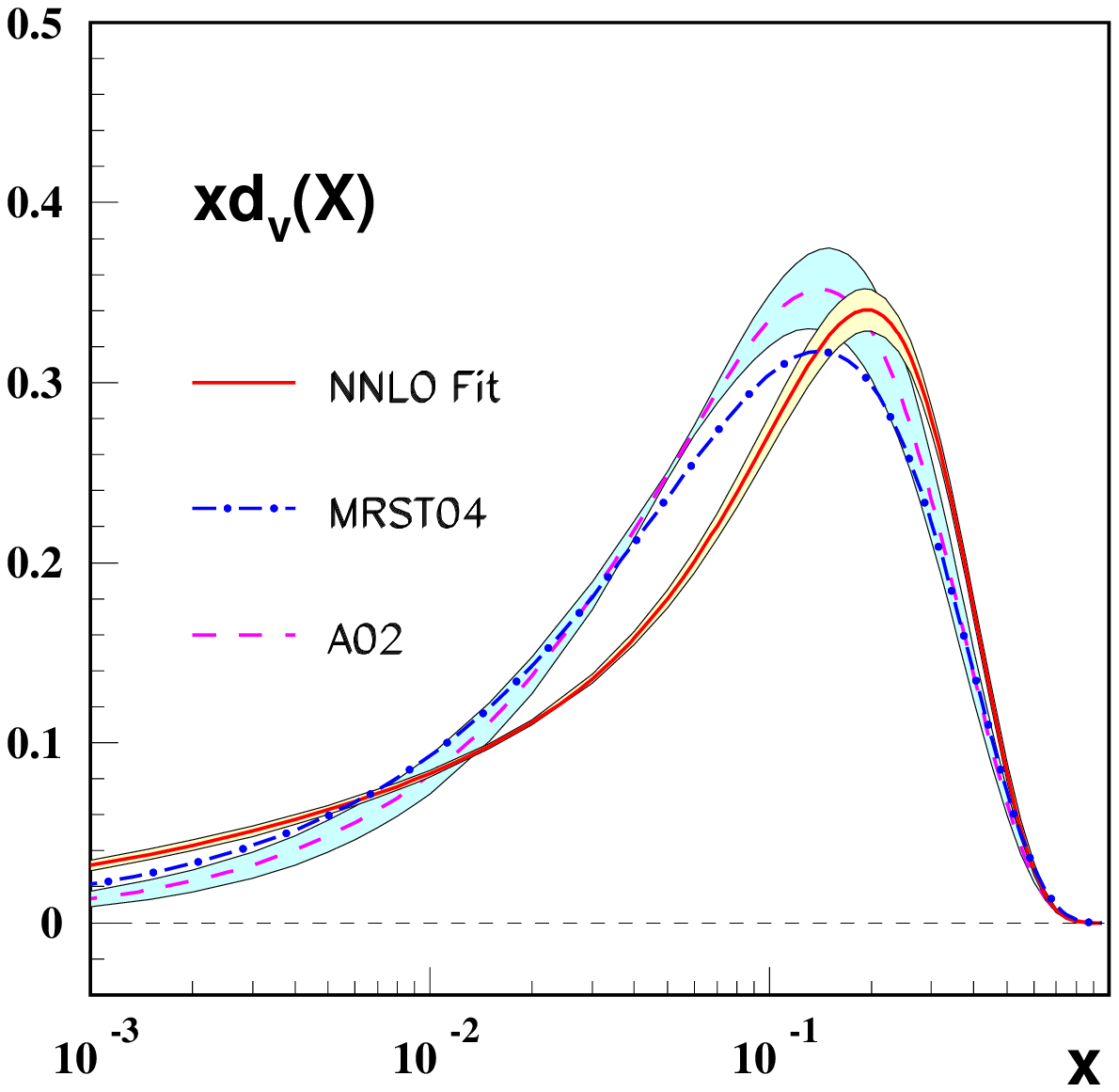}
\end{center}
\caption{The parton densities $xu_v$ (left) and $xd_v$ (right) at the input
scale $Q_0^2=4$ GeV$^2$ compared to results obtained by
MRST\protect\cite{Martin:2004ir} and Alekhin\protect\cite{Alekhin:2005gq}.
The shaded areas represent the fully correlated 1$\sigma$ error bands.
\label{fig:pdfs}}
\end{figure}

In Figure~\ref{fig:pdfs} we compare the parton distribution functions 
$xu_v(x)$  and $xd_v(x)$ at the reference scale $Q_0^2=4$ GeV$^2$ as 
extracted from our fit with the results obtained in other NNLO QCD fits.
\vspace{-.2cm}

\end{document}